\documentstyle[11pt,baaa-cas,twoside,epsf,psfig]{article}
\begin{document}
\vskip 1.0cm
\markboth{Fern\'andez Laj\'us et al.}{Campa\~na de observaci\'on de $\eta$~Carin\ae}
\pagestyle{myheadings}

%%%% DESCOMENTE LA LINEA QUE DESCRIBE EL CARACTER DE SU TRABAJO %%%%%%

%\parindent 0pt{INFORME INVITADO          -- INVITED REVIEW} 
%\parindent 0pt{ COMUNICACI\'ON DE TRABAJO -- CONTRIBUTED  PAPER } 
\parindent 0pt{INFORME DE TRABAJO -- WORK IN PROGRESS }
\vskip 0.3cm
\title{Campa\~na de observaci\'on de $\eta$~Carin\ae\ desde La Plata. Per\'{\i}odo 2003-2004} 

\author{
E.~Fern\'andez Laj\'us$^{1,2}$, 
C.~Fari\~na$^{1}$, 
R.~Gamen$^{3}$, 
C.~LLinares$^{1}$,
N.~Salerno$^{1}$, 
M.~Schwartz$^{1}$, 
L.~Simontacchi$^{1}$, 
A.~Torres$^{1,2}$,
V.~Niemela$^{1,4}$
}

\affil{
1- 
Facultad de Ciencias Astron\'omicas y Geof\'{\i}sicas - UNLP, Argentina.
2- Becario de CONICET, Argentina.\\
3- Departamento de F\'{\i}sica, Universidad de La Serena, Chile.\\
4- Miembro de la Carrera del Investigador Cient\'{\i}fico, CIC, Bs. As., Argentina.\\
%E-mail: eflajus@fcaglp.unlp.edu.ar
}

\begin{abstract} After the Observing Campaign of $\eta$~Carin\ae\ carried
out during 2003, we have continued the CCD observations 
using the 0.8 m Reflector telescope at La Plata Observatory, in the 
$BVRI$ bands.
Here, we present the results obtained since November 2003 until August 2004.
In addition, we present differential photometry of other objects included 
in the $\eta$~Car's frames, belonging to the open cluster Trumpler 16.
\end{abstract}

\begin{resumen} Luego de la campa\~na de observaci\'on de  $\eta$~Carin\ae\ 
llevada a cabo durante el 2003, hemos continuado las 
observaciones CCD con el telescopio Reflector de 0.8 m del Observatorio 
de La Plata, en las bandas $BVRI$.
Presentamos aqu\'{\i} los resultados obtenidos en el periodo noviembre de 
2003 a agosto de 2004.
Simult\'aneamente presentamos la fotometr\'{\i}a di\-fe\-ren\-cial de otros 
objetos inclu\'{\i}dos en las im\'agenes de $\eta$~Car, pertenecientes 
al c\'umulo abierto Trumpler 16.
\end{resumen}

\section{Introducci\'on}
$\eta$~Carin\ae\ es la Variable Luminosa Azul m\'as brillante del cielo,
y la concentraci\'on de estrellas OB en su entorno es com\'unmente denominada
regi\'on de Eta Carina. La regi\'on HII ionizada por estas estrellas OB,
llamada "Nebulosa de Carina", es una de las m\'as espectaculares de nuestra Galaxia.
La estrella $\eta$~Carin\ae\ est\'a envuelta en su propia nebulosa, 
el "Homunculus", producto de la eyecci\'on de masa durante una 
impresionante erupci\'on ocurrida a mediados del siglo XIX.

Entre enero y agosto del a\~no 2003 realizamos una campa\~na de observaci\'on de
fotometr\'{\i}a CCD \'optica de $\eta$~Carin\ae\
desde el Observatorio de La Plata (OALP)
perteneciente a la Facultad de Ciencias Astron\'omicas y Geof\'{\i}sicas, 
Universidad Nacional de La Plata (FCAG-UNLP), Argentina.
Las observaciones fueron
lle\-va\-das a cabo en el marco de una campa\~na internacional multifrecuencia
para monitorear a este objeto durante un m\'{\i}nimo de rayos X esperado
para el 2003.5, el cual efectivamente se produjo. Nuestras observaciones
\'opticas detectaron un m\'{\i}nimo una semana despu\'es del m\'{\i}nimo
en rayos X.
Estos resultados han sido publicados por Fern\'andez Laj\'us et al. (2003).

\begin{figure}
{
%{\hspace*{0.1cm} 
{\bf a} \psfig{figure=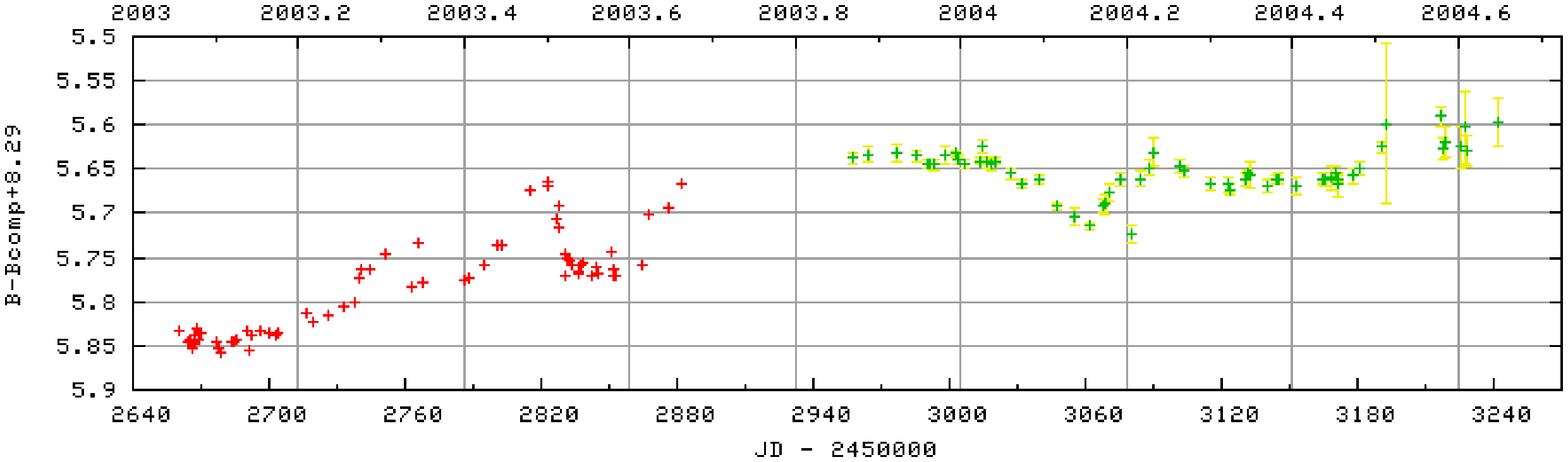,width=13.0cm,height=4.0cm,clip=,angle=-90}
%}{\hspace*{0.1cm} 
{\bf b} \psfig{figure=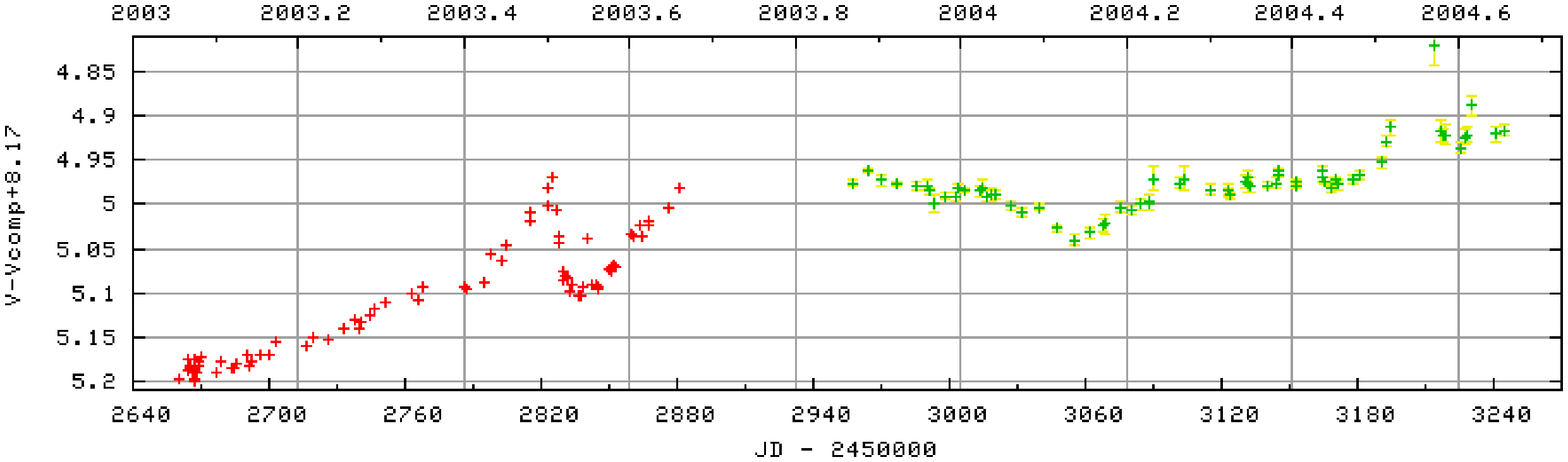,width=13.0cm,height=4.0cm,clip=,angle=-90}
%}{\hspace*{0.1cm} 
{\bf c} \psfig{figure=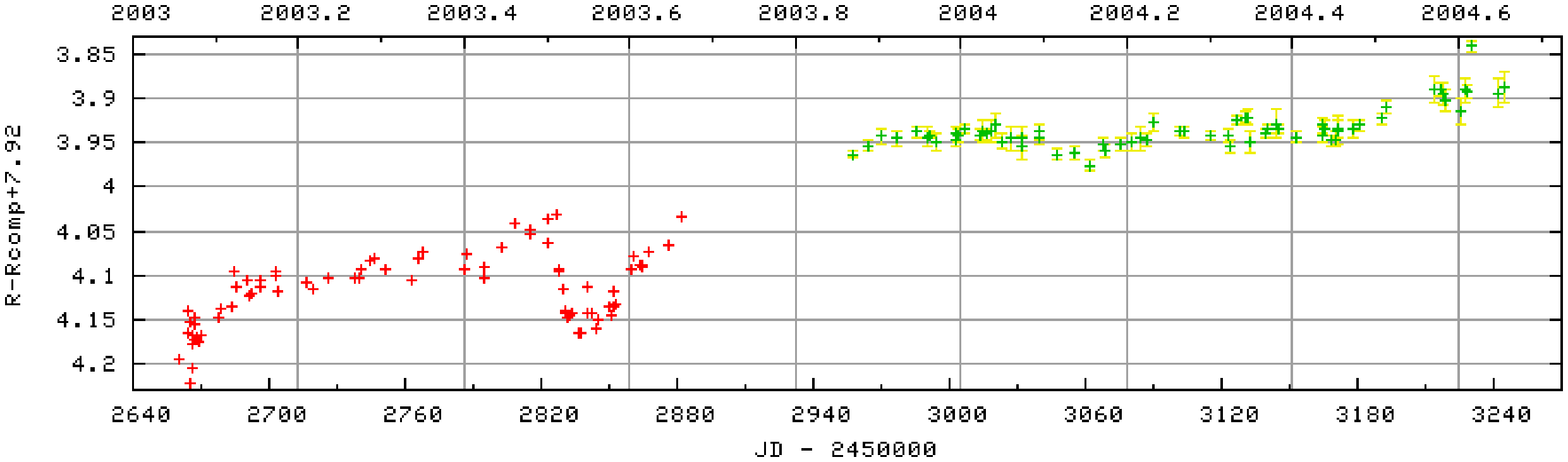,width=13.0cm,height=4.0cm,clip=,angle=-90}
%}{\hspace*{0.1cm} 
{\bf d} \psfig{figure=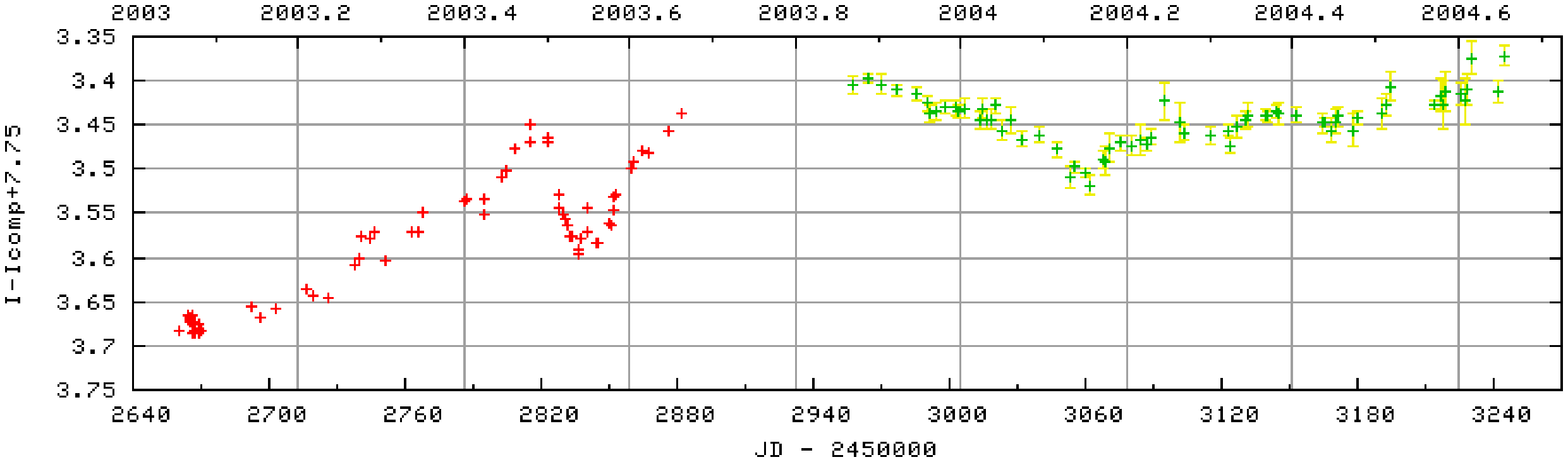,width=13.0cm,height=4.0cm,clip=,angle=-90}
%}
}
\caption{Curvas de luz $BVRI$ de $\eta$ Car resultantes de las observaciones
realizadas entre enero de 2003 y agosto de 2004.
Los puntos rojos corresponden a los datos obtenidos entre el
20 de Enero y el 29 de Agosto de 2003 (Fern\'andez Laj\'us et al. 2003)
y los verdes a los obtenidos entre el 14 de noviembre de 2003 y el
27 de agosto de 2004.}
\label{EtaCar}
\end{figure}

Desde noviembre de 2003 hasta agosto de 2004 hemos continuado
las observaciones de $\eta$~Carin\ae\, ya que se piensa que contiene
un sistema binario con un per\'{\i}odo de 5.5 a\~nos (Damineli et
al. 2000) y con\-si\-de\-ra\-mos de gran importancia realizar un seguimiento
fotom\'etrico durante todo el per\'{\i}odo orbital.

Presentamos aqu\'{\i} nuestros resultados m\'as recientes de la
fotometr\'{\i}a diferencial en las bandas $BVRI$ de Johnson-Cousins.
Mostramos, adem\'as, la fotometr\'{\i}a diferencial de otras dos 
estrellas en el mismo campo.

\begin{figure}
%\hbox
{
\hbox{\hspace*{1.3cm}
 \psfig{figure=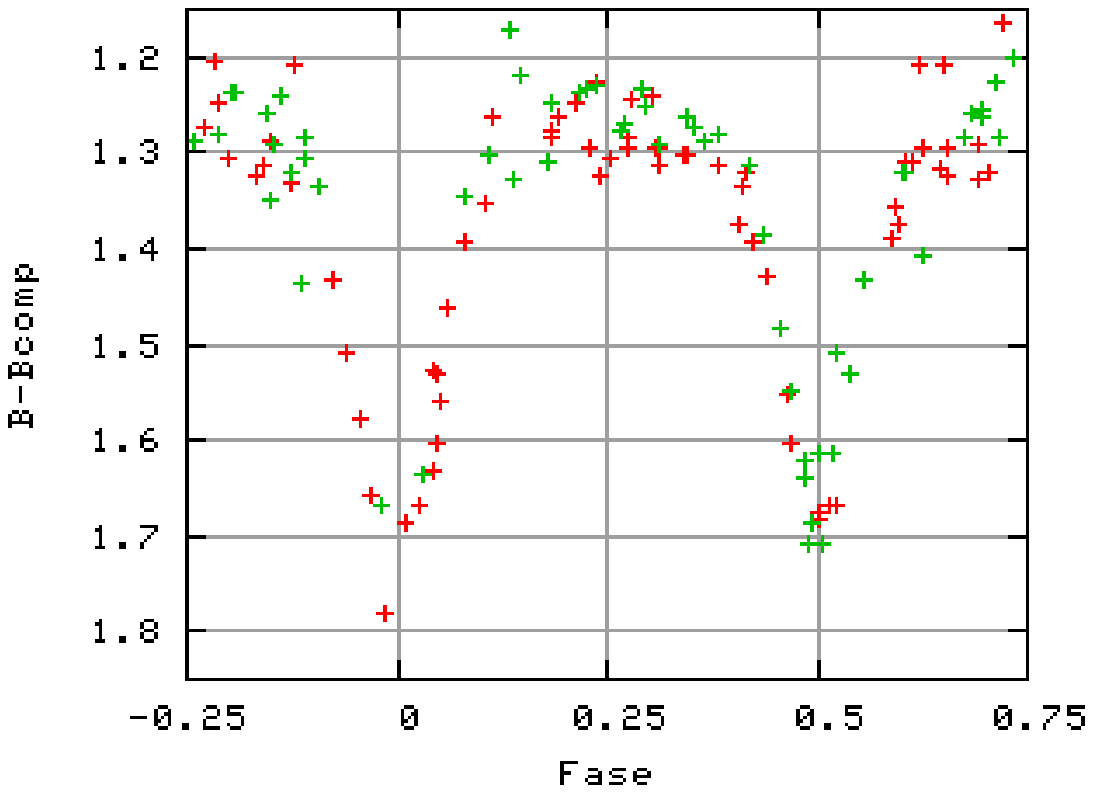,width=5.cm,height=3.3cm,clip=,angle=-00}
%\vskip 0.2cm
 \psfig{figure=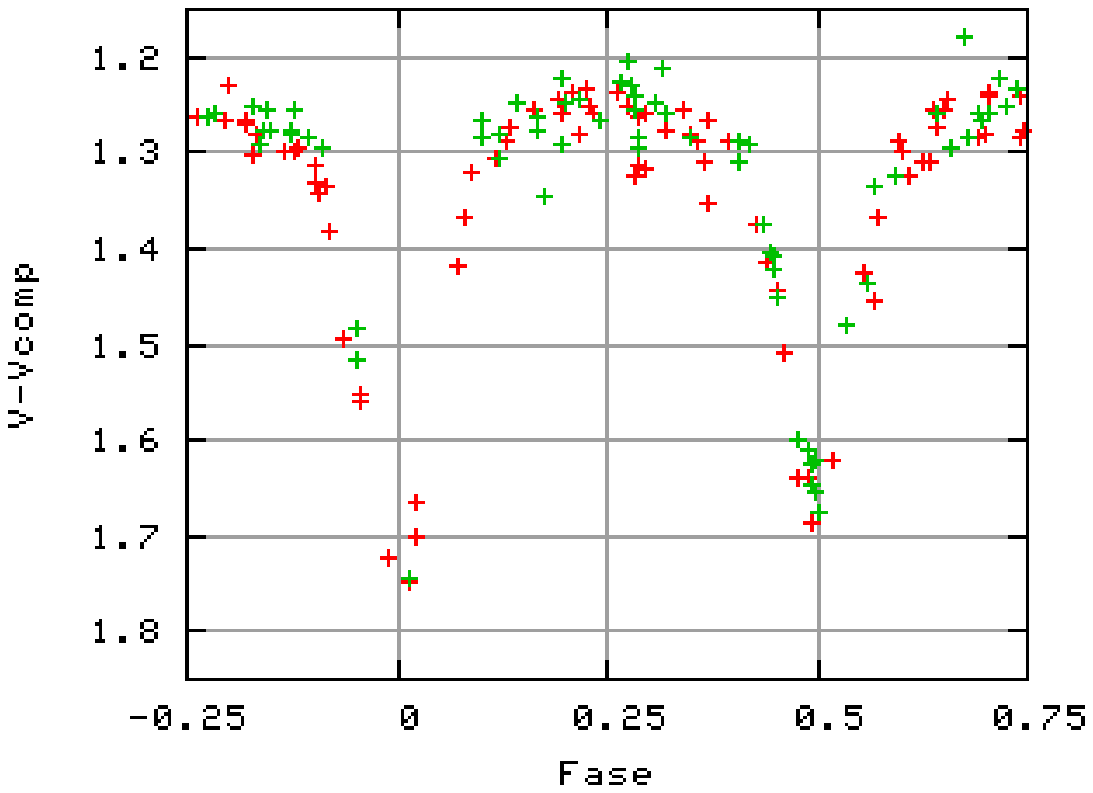,width=5.cm,height=3.3cm,clip=,angle=-00}
}
%\vskip 0.2cm
\hbox{\hspace*{1.3cm}
 \psfig{figure=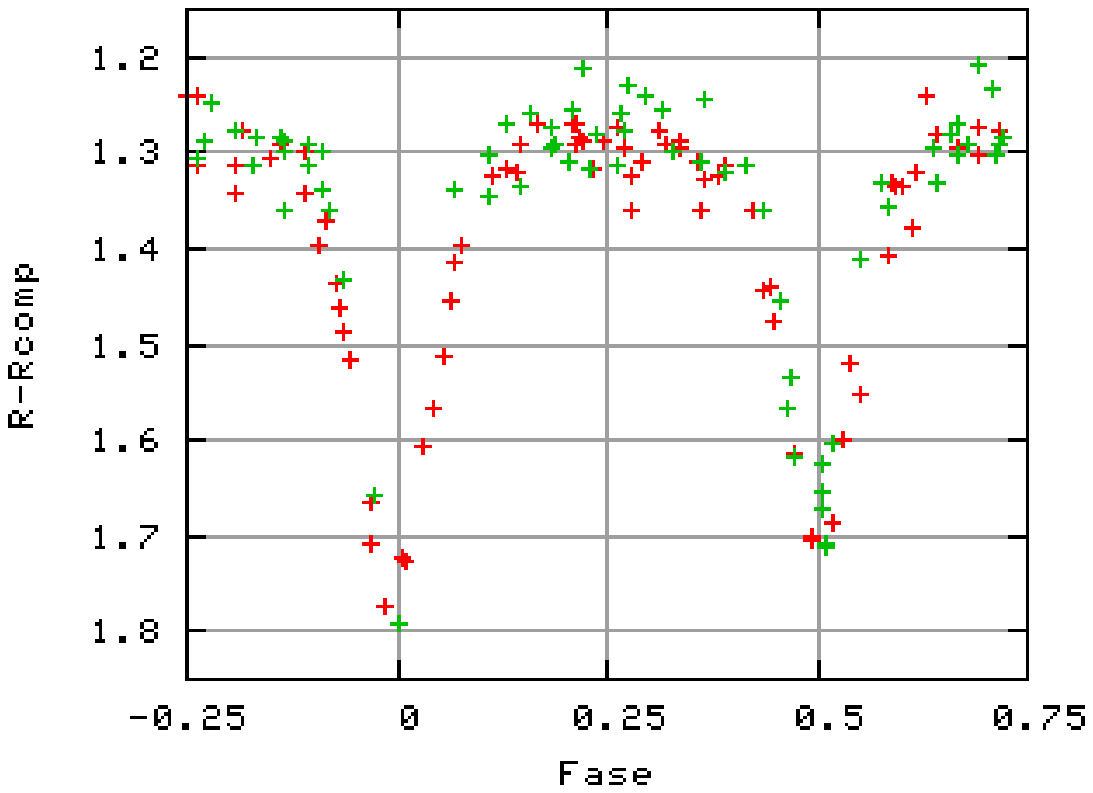,width=5.cm,height=3.3cm,clip=,angle=-00}
%\vskip 0.2cm
 \psfig{figure=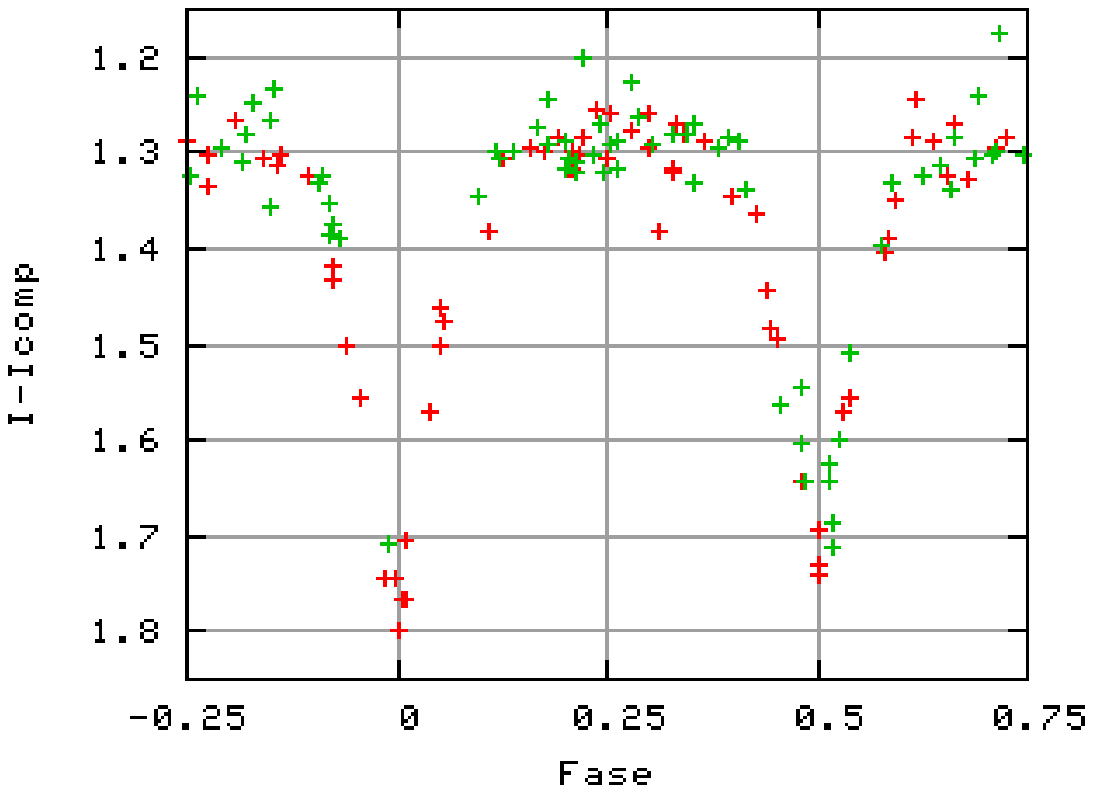,width=5.cm,height=3.3cm,clip=,angle=-00}
}
}
\caption{Curvas de luz $BVRI$ de CPD -59 2628 seg\'un la fase
orbital 2450456.8164+1.4693316E 
%d\'{\i}as 
(Freyhammer et al., 2001).
S\'{\i}mbolos idem Fig.~\ref{EtaCar}.}
\label{Tr16-1}
\end{figure}

\section {Observaciones}
Las observaciones fotom\'etricas fueron llevadas a cabo con el
telescopio Reflector de 0.80-m del OALP, con una c\'amara directa
Photometrics STAR I y un detector Thomson TH7883PS
de 384 x 576 pixels (23 $\mu m$/pixel). La configuraci\'on
instrumental resulta en im\'agenes de 1'54'' x 2'50'' de campo.
Se obtuvieron m\'as de 2600 im\'agenes en los filtros
$B V R I$ del sistema de Johnson-Cousins entre el 14 de noviembre
de 2003 y el 27 de agosto de 2004.

La fotometr\'{\i}a diferencial de $\eta$~Car
fue determinada usando HDE~303308 ($V$=8.15) como estrella de comparaci\'on.
Las magnitudes instrumentales de las estrellas fueron obtenidas
mediante fotometr\'{\i}a de apertura.
Dado que no es posible se\-pa\-rar el objeto central de $\eta$~Car
del Homunculus, se tomaron radios de apertura para la 
fotometr\'{\i}a de 75 pixels (22").
Para las otras estrellas del campo se utilizaron radios de apertura 
de 50 pix (15").
A partir de las magnitudes di\-fe\-ren\-cia\-les medidas en cada
imagen, se calcula un promedio pesado por los errores de cada medida
individual de una serie de im\'agenes ($n \sim 10$) en cada filtro.
La dispersi\'on del promedio es adoptada como el error de la medida
resultando errores t\'{\i}picos $\varepsilon_B = 0.008$, $\varepsilon_V = 0.005$, 
$\varepsilon_R = 0.007$ y $\varepsilon_I = 0.01 mag$.

\begin{figure}
{\hspace*{0.5cm}   \psfig{figure=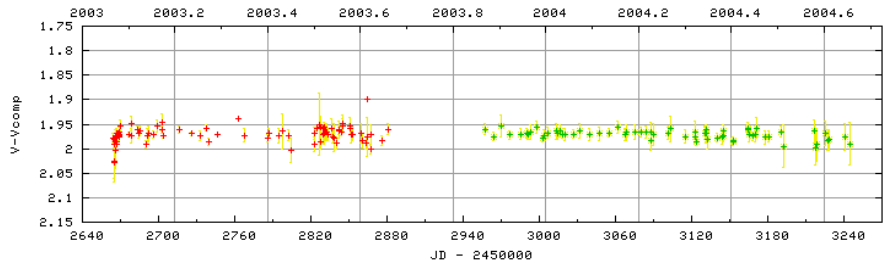,width=12.cm,angle=-90.0}}
\caption{Curva de luz $V$ de CPD -59 2627. S\'{\i}mbolos idem Fig.~\ref{EtaCar}.}
\label{Star4V}
\end{figure}

\section{Resultados}

La Figura~\ref{EtaCar} muestra las variaciones de luz de $\eta$ Car
observadas en La Plata a trav\'es de los filtros $B, V, R$ e $I$.
Los puntos rojos corresponden a las observaciones publicadas anteriormente.
En ellas se puede apreciar un m\'{\i}nimo en todos los filtros correspondiente
con el m\'{\i}nimo observado en rayos X en 2003.5.
Los puntos verdes corresponden 
al periodo de observaciones 2004
y en ellos se puede apreciar otro m\'{\i}nimo local en los 4 filtros
($\Delta_B \sim 0.07$, $\Delta_V \sim 0.05$, $\Delta_R \sim 0.02$, 
$\Delta_I \sim 0.07$) alrededor de DJ 2453061 para los filtros $B, R$ e $I$
y de DJ 2453055 para el $V$.
Este m\'{\i}nimo local se produce unos 220 d\'{\i}as despu\'es del
m\'{\i}nimo de julio de 2003, es decir en la fase 0.11 considerando
un per\'{\i}odo de 2022 d\'{\i}as
(P=5.536 a\~nos seg\'un 
Corcoran (2003) obtenidas a partir de las observaciones del RXTE).
Ning\'un m\'{\i}nimo an\'alogo a este \'ultimo fue observado en rayos X.

Puede notarse adem\'as que el sistema contin\'ua con un leve aumento de
brillo principalmente en las bandas $B, V$ y $R$.
Desde finales de 2003 ha superado casi
definitivamente la magnitud $V$=5, (excepto durante el m\'{\i}nimo 
mencionado en el p\'arrafo anterior) no alcanzada por la estrella desde 
1860, cuando $\eta$ Car se encontraba en su fase de decaimiento luego de la
erupci\'on de 1843. 

Como indicadores de la confiabilidad de nuestra fotometr\'{\i}a presentamos
observaciones de dos estrellas inclu\'{\i}das en el campo de $\eta$~Car.
En la Figura~\ref{Tr16-1} mostramos la curva de luz de la estrella binaria
eclipsante de doble espectro CPD-59 2628 (=Tr16-1, $V$=9.6). Nuestras observaciones
reproducen perfectamente las curvas publicadas por Freyhammer et al. (2001).

La Figura~\ref{Star4V} muestra las magnitudes diferenciales $V$ medidas en
la estrella CPD-59 2627 (=Tr16-3, $V$=10.2), la cual es 2 magnitudes m\'as 
d\'ebil que HDE 303308 
y m\'as de 5 magnitudes m\'as d\'ebil que $\eta$ Car.
Esta estrella resulta siempre subexpuesta en nuestras im\'agenes debido
a que los tiempos de exposici\'on est\'an optimizados para  $\eta$ Car.
Sin embargo, la dispersi\'on de los datos a lo largo de casi 2 a\~nos
es del orden de 0.025 mag. Esta estrella, dentro de los errores, no
parece haber variado su brillo durante nuestras observaciones.

%{\bf 
Todas las observaciones son publicadas y actualizadas permanentemente en: 
{\it
%\begin{center}
http://lilen.fcaglp.unlp.edu.ar/EtaCar
%\end{center}
}

\agradecimientos 
Agradecemos a las autoridades de la FCAyG-UNLP por facilitar los
recursos observacionales y al personal t\'ecnico del Observatorio
por sus aportes al mantenimiento del Telescopio Reflector.\\

\begin{referencias}

\reference Corcoran M. F. 2003,\\
http://lheawww.gsfc.nasa.gov/users/corcoran/eta\_car/2003.5/
\reference Damineli A., Kaufer A., Wolf B., Stahl O., Lopes D., de Ara\'ujo F., 2000, ApJ, 528, L101.
%\reference Feinstein A., Marraco H., \& Muzzio J.C., 1973, A\&AS, 12, 331.
\reference Fern\'andez Laj\'us E., Gamen R., Schwartz M., Salerno N., Llinares C.;
Fari\~na C., Amor\'{\i}n R., Niemela V., 2003, IBVS, 5477, 1.
\reference Freyhammer L., Clausen J., Arentoft T., \& Sterken C., 2001, A\&A, 369, 561.

\end{referencias}

\end{document}